\documentclass[12pt,a4paper]{article}
\usepackage{jcappub}
\usepackage[normalem]{ulem}
\usepackage{color}

\usepackage{euscript,amssymb,amsmath}
\usepackage{amsfonts,latexsym}
\usepackage{amsfonts,amsmath,amssymb,amsfonts,latexsym}
\usepackage[utf8]{inputenc}

\newcommand{\const}{\mbox{\rm const}\,}

\usepackage{graphicx}
\usepackage{epsfig}
\usepackage{color}

\newcommand{\be}[1]{\begin{equation}\label{#1}}
\newcommand{\ee}{\end{equation}}
\newcommand{\ba}[1]{\begin{eqnarray}\label{#1}}
\newcommand{\ea}{\end{eqnarray}}
\newcommand{\rf}[1]{(\ref{#1})}

\def\H{{\mathcal H}}      

\newcommand{\al}{\alpha}
\newcommand{\bt}{\beta}

\newcommand{\mf}[1]{\mathbf{#1}}

\begin{document}
	
\title{Effect of medium on fundamental interactions in gravity and condensed matter}

\author[a,b,c]{Alexander Zhuk}
\author[a,d,1]{Valerii Shulga}\note{corresponding author}

\affiliation[a]{International Center of Future Science of the Jilin University, \\2699 Qianjin St., 130012, Changchun City, China\\}

\affiliation[b]{Astronomical Observatory, Odessa National University,\\ Dvoryanskaya st. 2, Odessa 65082, Ukraine\\}

\affiliation[c]{Center for Advanced Systems Understanding (CASUS), \\Untermarkt 20, 02826 Görlitz, Germany\\} 

\affiliation[d]{Institute of Radio Astronomy of National Academy of Sciences of Ukraine, \\4 Mystetstv str., 61002 Kharkiv, Ukraine\\}

\emailAdd{ai.zhuk2@gmail.com}, 
\emailAdd{shulga@rian.kharkov.ua}

\abstract{Recently, it was shown that the gravitational field undergoes exponential cutoff at large cosmological scales due to the presence of background matter. In this article, 
we demonstrate that there is a close mathematical analogy between this effect and the behavior of the magnetic field induced by a solenoid placed in a superconductor.

\
	
}

\maketitle

\flushbottom

\section{Introduction}

It seems quite natural that the presence of the medium influences the propagation of fundamental interactions. The simplest example is the Debye screening of the electric field of an individual particle in a plasma by particles of opposite sign. Here, the potential produced by an external point charge has the form of the Yukawa potential (but not the Coulomb one) with the Debye screening length (see, e.g., \cite{Coulomb}). A similar screening mechanism of the electron charge due to vacuum polarization takes place in quantum electrodynamics (see, e.g., \cite{QED}). The Anderson-Higgs mechanism is another example of the influence of the medium on fundamental interactions, which are carried by gauge fields. In this case, after symmetry breaking, the Higgs vacuum field acts as a medium \cite{Higgs1,Higgs2,Higgs3}. As a result of interaction with this medium, the initially massless gauge fields gain mass \cite{Linde}. It is also known that medium in the form of the superconductor affects the electromagnetic interaction. For example, external magnetic field undergoes the exponential cutoff inside the superconductor due to the Meissner effect (see, e.g., \cite{Svist}).

The examples above did not concern the gravitational interaction between massive bodies. 
It is known that in a vacuum in the weak field limit the gravitational potential satisfies the Poisson equation and has the form of Newton's potential \cite{LL}. From a naive point of view, since all masses have the same sign and are attracted to each other, one should hardly expect a screening of the gravitational interaction, as, for example, for electric charges in a plasma. However,  it was demonstrated recently \cite{Ein1,EKZ1,EKZ2} that medium in the case of gravity also plays important role. It was shown that, due to the interaction of the gravitational potential with background matter, there is an exponential cutoff of the gravitational interaction at large cosmological scales. In section 2 we reproduce this result. For many, this result turned out to be rather unexpected. Therefore, in this paper, in section 3, we present a close mathematical analogue of this phenomenon by the example of the magnetic field induced by a solenoid placed in a superconductor.

\

\section{Screening of the gravitational interaction in cosmology}

\setcounter{equation}{0}

We consider the Universe containing the cosmological constant $\Lambda$ and filled with
discrete point-like gravitating sources 
(galaxies and the group of galaxies)
with comoving mass density
\be{2.1}
\rho = \sum_n \rho_n =  \sum_n m_n \, \delta(\mf{r}-\mf{r}_n) \, ,
\ee
where $\mf{r}=(x^1,x^2,x^3)$ is comoving distance. This is our medium. Such matter has a dust-like equation of state and the average energy density  $\bar\varepsilon = \bar{\rho}c^2/a^3$ where comoving averaged mass density $\bar\rho=\mathrm{const}$, $c$ is the speed of light and $a$ is the conformal factor.
The corresponding background metric is described  by Friedmann-Lema$\mathrm{\hat{\i}}$tre-Robertson-Walker (FLRW) one. 

The discrete inhomogeneities perturb the FLRW metric \cite{Mukhanov,Gorbunov:2011zzc}:
\be{2.2} ds^2 = a^2 \big[ (1+2\Phi) d\eta^2  - (1-2\Phi) \delta_{\al\bt} \, dx^\al dx^\bt\,\big]\, , \ee
where we restrict ourselves to scalar perturbations in conformal Newtonian gauge. Scalar function $\Phi(\eta,\mf{r})$ is the gravitational potential  created at the point with the radius-vector $\mf{r}$ by all gravitating masses in the Universe \cite{LL}.   The perturbed Einstein equations are \cite{Mukhanov,Gorbunov:2011zzc}:
\be{2.3}
\Delta\Phi-3\H\left(\Phi'+\H\Phi\right)=\frac12\kappa a^2\delta\varepsilon ,
\ee
\be{2.4}
\Phi'+\H\Phi=-\frac12 \kappa a^2\bar\varepsilon v\, ,
\ee
\be{2.5}
\Phi''+3\H\Phi'+\left(2\H'-\H^2\right)\Phi=0\, ,
\ee
where $\Delta\equiv \delta^{\alpha\beta}\partial_{\alpha}\partial_{\beta}$ is the Laplace operator, the prime denotes the conformal time $\eta$ derivative, $\H\equiv (da/d\eta)/a =(a/c)H$ and $H\equiv (da/dt)/a$ is the Hubble parameter, $v(\eta,\textbf{r})$ is the peculiar velocity potential and $\kappa \equiv 8\pi G_{\!N}/c^4$, where $G_{\!N}$ is the gravitational constant. The energy density fluctuation reads \cite{EZflow,EZremarks}:
\be{2.6}
\delta\varepsilon = \frac{c^2}{a^3}\delta\rho + \frac{3\bar\rho c^2}{a^3}\Phi\, ,
\ee
where $\delta\rho (\eta,\mf{r})\equiv \rho-\bar\rho$ is the fluctuation of the mass density \rf{2.1} around its constant average value $\bar\rho$.

Eq. \rf{2.4} demonstrates that the peculiar velocities affect the gravitational potential. If we neglect this influence (i.e. $\Phi'+\H\phi =0$), then equation \rf{2.3} takes the form
\be{2.7}
\Delta\Phi -\frac{a^2}{\lambda^2}\Phi = \frac{\kappa c^2}{2a}\delta\rho\, ,
\ee
where the screening length
\be{2.8}
\lambda\equiv \sqrt{\frac{2a^3}{3\kappa\bar\rho c^2}}\, .
\ee
With the help of the transformation (to remove the $\bar\rho$ contribution on the RHS of Eq. \rf{2.7})
\be{2.9}
\phi =c^2 a\Phi -\frac{4\pi G_N\bar\rho}{a^2}\lambda^2 = c^2 a\Phi - \frac{c^2 a}{3}
\ee
Eq. \rf{2.7} is reduced to
\be{2.10}
\Delta\phi -\frac{a^2}{\lambda^2}\phi = 4\pi G_N \rho\, .
\ee
For the mass density \rf{2.1}, we can easily solve this Helmholtz equation, and applying transformation \rf{2.9} obtain:
\be{2.11}
\Phi = \frac13 -\frac{\kappa c^2}{8\pi a}\sum_n \frac{m_n}{|\mf{r}-\mf{r}_n|}\exp\left(-\frac{a|\mf{r}-\mf{r}_n|}{\lambda}\right)\, .
\ee
It is worth noting that the physical distance is $R=ar$.
The term 1/3 (which is due to  $\bar\rho$ in $\delta \rho$) plays an important role since only with this term the averaged over all volume value of the gravitational potential $\bar\Phi$ is equal to zero as it should be for fluctuations \cite{Ein1}. 

In Eq. \rf{2.11}, we neglect the peculiar velocities of the inhomogeneities. However, they also play an important role \cite{pecvel1,pecvel2} and must be taken into account. For the considered model, as was shown in  \cite{pecvel1}, it is sufficient in \rf{2.7}, \rf{2.9}-\rf{2.11} to replace  $\lambda$ with $\lambda_{\mathtt{eff}}$ and additionally in \rf{2.11}: $1/3 \to 1/3 (\lambda_{\mathtt{eff}}/\lambda)^2$ where
\be{2.12}
\lambda_{\mathtt{eff}}= \sqrt{\frac{c^2 a^2 H}{3}\int\frac{da}{a^3H^3}}\, .
\ee
To get this result, we should take into consideration Eq. \rf{2.5}.  This screening length (as well as $\lambda$) depends on time. For example, for the standard $\Lambda$CDM model at present time $\lambda_{\mathtt{eff}} =2.57$ Gpc \cite{pecvel1}.

Therefore, the gravitational potential $\Phi$ satisfies the Helmholtz equation, not the Poisson equation. This is due to the interaction of the gravitational potential with the medium. We can see it directly from Eq. \rf{2.6} where the term $\sim \bar\rho\Phi$ describe this interaction. Due to the peculiar velocity, Eq. \rf{2.3} also acquires an additional term proportional to $\Phi$ \cite{pecvel1}. If the medium is absent that corresponds to the limit  $\bar\rho\to 0, \,  v\to 0$, then the screening lengths $\lambda$ and $\lambda_{\mathtt{eff}}$ tends to infinity, and the Yukawa potentials in \rf{2.11} are reduced to the Newton's ones without screening of the gravitational interaction.


\section{Solenoid in a superconductor. Screening of the induced magnetic field.}

\

In this section, in order to present the mathematical analog of the screening effect described above, we render some of equations of the paper \cite{solenoid} in a form suitable for our purpose.
Following this paper,
we consider a thin solenoid placed in a superconductor.  Thin means that the diameter of the solenoid is much smaller than the magnetic field penetration length $\lambda_{\mathtt{m}}$.
It is well known that the magnetic field of the solenoid $\mf{B}_{\mathtt{sol}}$  is absent from the outside it, but the vector potential $\mf{A}_{\mathtt{sol}}$  is not equal to zero. The interaction of this potential with the superconducting medium induces a current $\mf{J}_{\mathtt{sc}}$, which, in turn, leads to the appearance of an induced magnetic field $\mf{B}_{\mathtt{sc}}$. Thus, the Maxwell equation has the form{\footnote{In this section, we use the system of units adopted in book \cite{Svist}.}}
\be{3.1}
\mathtt{curl}\mf{B}_{\mathtt{tot}}=
\mathtt{curl}\left(\mf{B}_{\mathtt{sc}}+ \mf{B}_{\mathtt{sol}}\right)
= \mf{J}_{\mathtt{sc}} + \mf{J}_{\mathtt{sol}}\, .
\ee
Since outside the solenoid $\mf{B}_{\mathtt{sol}},\mf{J}_{\mathtt{sol}}=0$, we get
\be{3.2}
\mathtt{curl}\mf{B}_{\mathtt{sc}} 
= \mf{J}_{\mathtt{sc}} \, ,
\ee
where in the London limit the superconducting current density is \cite{Svist,solenoid}
\be{3.3}
\mf{J}_{\mathtt{sc}} = -\frac{1}{\lambda^2_{\mathtt{m}}}\left(\frac{1}{q}\nabla \theta +\mf{A}_{\mathtt{tot}}\right)\, .
\ee
Here, $\mf{A}_{\mathtt{tot}}=\mf{A}_{\mathtt{sc}}+\mf{A}_{\mathtt{sol}}$, $\theta$ is the phase of the order parameter and the magnetic field penetration length 
\be{3.4}
\lambda_{\mathtt{m}}=\frac{1}{q\sqrt{n_{\mathtt{s}}}}\, ,
\ee
whrere $n_{\mathtt{s}}$ is the superfluid density, parameter $q$ defines the superconducting flux quanta (see, e.g., Eq. \rf{3.7} below) and in the real superconductor $q=2e/(\hbar c)$ \cite{Svist}.
The absence of a superconducting medium corresponds to the limit $n_{\mathtt{s}}\to 0 \Rightarrow \lambda_{\mathtt{m}} \to \infty$. Expression \rf{3.4} is an analogue of cosmological formula \rf{2.8} (and, accordingly, formula \rf{2.12}). In Eq. \rf{3.3} the term $\lambda^{-2}_{\mathtt{m}}\mf{A}_{\mathtt{sol}}\sim n_{\mathtt{s}}\mf{A}_{\mathtt{sol}}$ describes the interaction of the solenoid magnetic field with the superconducting medium just as the term $\sim \bar\rho\Phi$ on the RHS of  Eq.  \rf{2.6} describes the interactions of the gravitational potential with the cosmological medium.

Now, applying curl operation to both sides of \rf{3.3}, we obtain
\be{3.5}
\mf{B}_{\mathtt{sc}}
- \lambda^{2}_{\mathtt{m}}\Delta \mf{B}_{\mathtt{sc}} =0 \, ,
\ee
where we took into account that outside of the solenoid $\mathtt{curl}\nabla\theta=0$
and $\mf{B}_{\mathtt{sol}}=0$.
$\Delta$ is the Laplace operator in flat space. To solve this equation, we need to define the boundary conditions. 
Let the solenoid be extended along the $z$-axis. Obviously, due to the cylindrical symmetry the induced magnetic field inside the superconductor is also parallel to the $z$-axis: $\mf{B}_{\mathtt{sc}} (r) = {B}_{\mathtt{sc}} (r) \hat{z}$, where $\hat{z}$ is the unit vector along $z$-axis. In cylindric coordinates, $\mf{r}$ is the radius-vector in the $xy$-plane (it is worth noting that in the previous section $\mf{r}$ denotes the comoving three-dimensional radius-vector). At  distances $r>>\lambda_{\mathtt{m}}$, the superconducting current goes to zero: $\mf{J}_{\mathtt{sc}} \to 0$. Therefore, at this distances Eq. \rf{3.3} reads
\be{3.6}
 \mf{A}_{\mathtt{tot}} = -\frac{1}{q}\nabla \theta\, .
\ee
Integrating both sides of this equation over an area inside the contour $r=\const$, and performing the Stokes area-to-contour transformation for the RHS, we find
\be{3.7}
\Phi_{\mathtt{tot}} = -\frac{2\pi}{q}N\equiv -\Phi_0 N\, ,\quad N=0,1,2,\ldots\, ,
\ee
where $\Phi_{\mathtt{tot}}=\Phi_{\mathtt{sc}}+\Phi_{\mathtt{sol}}$ is the total magnetic flux consisting of the sum of the magnetic fluxes of the induced magnetic field and the magnetic field inside the solenoid. $\Phi_0$ is the superconducting flux quanta. Therefore,
\be{3.8}
\Phi_{\mathtt{sc}} = \Phi_{\mathtt{tot}}-\Phi_{\mathtt{sol}}\, .
\ee
This is our boundary condition. We can include it directly into Eq. \rf{3.5}:
\be{3.9}
B_{\mathtt{sc}}
- \lambda^{2}_{\mathtt{m}}\Delta B_{\mathtt{sc}} = \Phi_{\mathtt{sc}} \delta(\mf{r})\, ,
\ee
where we took into account 2D cylindrical symmetry of the problem and, consequently, $\Delta$ is a radial Laplace operator. Obviously, integrating this equation over an area inside the contour $r=\const$ we arrive at  identity. Eq. \rf{3.9} is the Helmholtz one (similar to Eq. \rf{2.10}), and has the decreasing solution
\be{3.10}
B_{\mathtt{sc}}=\frac{\Phi_{\mathtt{sc}}}{2\pi\lambda^2_{\mathtt{m}}}
K_0\left(\frac{r}{\lambda_{\mathtt{m}}}\right)\, ,
\ee
where $K_0$ is the modified Bessel function.  The induced magnetic field behaves asymptotically as follows:
\be{3.11}
B_{\mathtt{sc}} (r\to 0)\sim -\ln (r)\, ,\quad
B_{\mathtt{sc}} (r\to \infty) \sim \frac{1}{\sqrt{r}} \exp(-r/\lambda_{\mathtt{m}})\, 
\ee
This behavior reflects  the cylindrical symmetry of the model. For example, Yukawa's potential has been transformed:
$(1/r)\exp(-r/\lambda_{\mathtt{m}}) \to
(1/\sqrt{r})\exp(-r/\lambda_{\mathtt{m}})$.
As expected, the screening length coincides with the magnetic field penetration length $\lambda_{\mathtt{m}}$. Formula \rf{3.11} is 2D analog of Eq. \rf{2.11}.


\


\section{Conclusion}

\

In this paper, we have touched upon the problem of the influence of the medium on fundamental interactions. First, on the basis of articles \cite{Ein1,EKZ1,EKZ2}, we showed that as a result of the interaction of the gravitational field with the cosmological medium, the gravitational potential is subject to exponential screening on large cosmological scales. Then, following the model considered in paper \cite{solenoid}, we have traced a close analogy between the interaction of the gravitational field with the cosmological medium and the interaction of the magnetic field of a solenoid with a superconducting medium.
As a result of this interaction, the induced magnetic field in the superconductor undergoes exponential screening at distances exceeding the magnetic field penetration length.


\







\section*{Acknowledgements}

The authors are grateful to Boris Svistunov for fruitful discussions and valuable comments.




\end{document}